\documentclass[referee,onecolumn,extra]{gji}

\usepackage{psfrag}
\usepackage{amsmath}
\usepackage{amssymb}
\usepackage{def}
\usepackage[dvips]{epsfig}
\usepackage{graphicx}
\usepackage[OT1]{fontenc}

\renewcommand{\cite}[1]{\citet{#1}}

\begin{document}

\author[E. Chaljub and B. Valette]{%
Emmanuel Chaljub$^1$ and Bernard Valette$^2$\\
$^{1}$LGIT, CNRS, BP 53, 38041 Grenoble Cedex 9, France\\
$^{2}$LGIT, IRD, Universit\'e de Savoie, 73376 Le Bourget-du-Lac Cedex, France
}

\title[Modeling wave propagation in a self-gravitating Earth]{
Spectral element modeling of three dimensional wave propagation in a self-gravitating Earth
with an arbitrarily stratified outer core
}

\maketitle


\begin{abstract}
This paper deals with the spectral element modeling of seismic wave propagation at the global scale.
Two aspects relevant to low-frequency studies are particularly emphasized.
First, the method is generalized beyond the Cowling approximation in order to fully account for
the effects of self-gravitation.
In particular, the perturbation of the gravity field outside the Earth is handled by a projection of the 
spectral element solution onto the basis of spherical harmonics.
Second, we propose a new formulation inside the fluid which allows to account for an 
arbitrary density stratification.
It is based upon a decomposition of the displacement into two scalar potentials, and results in
a fully explicit fluid-solid coupling strategy.
The implementation of the method is carefully detailed and its accuracy is demonstrated through a series
of benchmark tests.
\end{abstract}

\renewcommand{\sectionmark}[1]{}
\section{Introduction}

It has been recently established by several authors 
(\cite{Cha00}, \cite{KomTro02a,KomTro02b}, \cite{Capetal03}, \cite{ChaCapVil03})
that the spectral element method (SEM) provides an efficient solution to the
issue of computing synthetic seismograms in three dimensional (3D) models of the Earth.
Whereas most of current spectral element studies aim at pushing calculations 
toward high frequencies, 
where the methods traditionally used at the global scale reach their limits,
this paper focuses on some physical effects that are critical for the lower 
part of the seismic frequency band:
(i) the full treatment of self-gravitation and
(ii) the ability to take into account any density stratification in the fluid regions of the Earth.

The first novelty of this paper stands in the incorporation of self-gravitation, the effect of which
is important for seismic and gravimetric observations with periods larger than 100 s.
All the previously mentioned studies based upon the SEM accounted for the effects of gravity within the 
Cowling approximation \citep{Cow41}, \ie by neglecting the perturbation of the gravity field by
seismic waves.
The main reason for making this assumption lies in the intrinsic difficulty of the problem.
Considering the full effects of self-gravitation requires, indeed, to solve Poisson's equation 
for the perturbed gravitational potential which is defined over the whole space.
Unlike spherical harmonics approaches, the use of a grid-based method such as the SEM does not provide
a natural framework for the resolution of the exterior problem.
Grid-based approximations in unbounded domains proceed first by restricting the computational domain, then
by imposing an appropriate condition on the truncating boundary.
Different methods arise depending on whether the artificial boundary condition (ABC) is local or not. 
Methods based upon a local ABC have the advantage of being computationally inexpensive and 
valid for arbitrary geometries.
An example of such methods is the infinite element method (\eg \cite{Bet92}, \cite{GerDem96}),
in which the behaviour of the exterior solution is enforced in the radial direction.
The second class of methods, based upon a non-local ABC, are not as general since they usually require 
the knowledge of an analytical, or semi-analytical, solution to the exterior problem. 
As a consequence they have very attractive properties regarding their accuracy while 
being restricted to simple (usually spherical) geometries.
The non-local ABC can be implemented into the finite element method within the rigorous
framework of a Dirichlet-to-Neumann ($DtN$) operator (\eg \cite{Giv92}).
This is the approach that we retain here.
The $DtN$ operator that suits our problem relies on the spherical harmonic decomposition of 
the solution of Laplace's equation outside the Earth.
Unlike the one introduced by \cite{Capetal03} to couple a time-dependent spectral element 
calculation to a modal solution in the frequency domain, our $DtN$ operator is much simpler to derive 
because it is applied to a static problem.
The spectral element discretization of the Poisson-Laplace equation yields a symmetric algebraic system
which has to be inverted at each time step to obtain the perturbation of the gravitational potential. 
In practice, this is done by iterating a conjugate gradient method, 
the preconditioning of which is critical to carry out routine calculations.

The other aspect we consider in great detail is the treatment of the fluid part of the 
Earth's core. 
A parameter which is of particular importance with regard to core dynamics 
is the squared \BV \ frequency $N^2$ that characterizes the local response of the fluid to 
perturbations in density.
To first order, the core can be considered as neutrally stratified, \ie $N^2=0$, 
because a neutral buoyancy is expected in the bulk of a region subject to vigorous convection.
However, there is seismological evidence for a negative $N^2$ at the top of the core and a positive
$N^2$ at its bottom, with absolute values that can reach $10^{-7} \ \rad^2\cdot\s^{-2}$
(\cite{Masters79}, Valette \& Lesage (unpublished)).
For the sake of generality, our description of the core's structure will
make no assumption on the profile of the buoyancy frequency.
To this end, we introduce a two-potential formulation of the wave equation in the fluid
that generalizes the neutral buoyancy formulation of \cite{KomTro02b} and \cite{ChaCapVil03}.
Contrary to these studies, that considered the velocity potential in the fluid, our decomposition
is applied to the displacement field in order to obtain natural solid-fluid boundary
conditions for the perturbed gravitational potential.
An attractive consequence of this choice is to yield a fully explicit solid-fluid 
coupling strategy, as opposed to the studies mentioned above.
Note finally that our formulation is close to the two-potential description
proposed by \cite{WuRoc90} in the context of core dynamics studies, which is optimal with respect to the
number of unknowns in the fluid regions. 

The remainder of the paper is organized as follows.
In section \ref{sec: equations}, we recall the equations of motion in a self-gravitating Earth 
in their strong and weak form, successively.
We introduce in particular the two-potential decomposition of the displacement field in the fluid regions
and we define the $DtN$ operator that permits to handle the equations within a finite domain.
In section \ref{sec: specfem}, we recall the principles of the spectral element approximation 
in space and we make a detailed presentation of our explicit time marching algorithm.
Finally, numerical results are shown in section \ref{sec: results} for a set of spherically symmetric 
models that validate the implementation of the method.

\section{Wave equation in a self-gravitating Earth}\label{sec: equations}
%
In this section we recall the strong and weak forms of the wave equation, which is 
obtained through a first order Lagrangian perturbation around a
non-rotating, hydrostatically pre-stressed, state of equilibrium.
Throughout the paper, the Earth is denoted by \earth \ and its outer boundary by \surf.
The solid (resp.~fluid) parts of \earth \ are referred to as $\VS$ (resp.~$\VF$), and the
set of all solid-fluid interfaces is denoted $\SSF$.
Whenever topography or ellipticity is considered on \surf , $\earthb$ will denote a ball of radius $b$
that contains the aspherical Earth (\ie, $\earth \subset \earthb$) and
$\surfb$ will stand for its spherical boundary ($\surfb = \partial\earthb$).

\subsection{Strong form}
%
Solving the wave equation within the previous assumptions consists in finding the Lagrangian
perturbation of the displacement, $\bu$, such that:
\begin{eqnarray}
\ddot{\bu} \ + \ \bcA(\bu) 
& = &  \frac{1}{\rho}\, \bbf \ , \label{eq: motion_sol_a} \\
 \rho\, \bcA(\bu) & = & 
-\nabla\cdot \bT(\bu) \ - \ 
\nabla \left( \rho \bu \cdot \bg \right) \ + \ 
\left\{\divrhou\right\} \, \bg \ + \ 
\rho\nabla\psi  \ ,
\label{eq: motion_sol_b} 
\end{eqnarray}
where $\bcA$ is the elastic-gravitational operator, $\bT(\bu)$ is the Lagrangian incremental stress tensor, 
$\rho$ is density, $\bg$ is the acceleration due to gravity, 
$\psi$ is the Eulerian perturbation of the gravitational potential, 
also known as the mass redistribution potential (MRP), and $\bbf$ is the forcing term.
As usual, a dot over a symbol implies time derivation and $\nabla\btau$ (resp.~$\nabla\cdot\btau$) 
stands for the gradient (resp.~the divergence) of a given tensor field $\btau$.

In the (inviscid) fluid regions the stress tensor takes the form:
\begin{equation}\label{def: T_liq}
\bT(\bu) \ = \ \rho c^2 \, \divu \ \bI \ ,
\end{equation}
where $c$ is the speed of sound and $\bI$ denotes the second-order identity tensor. 
Neglecting any source term in the fluid, the wave equation can then be rewritten as:
\begin{equation}\label{eq: motion_liq_u}
\ddot{\bu} \ = \ -\bcA(\bu) \ = \ 
\nabla \left[
c^2\divu \ + \ \bu\cdot\bg \ - \ \psi
\right] 
\ + \ c^2 \, \divu \ \bs
\ ,
\end{equation}
where $\bs$ is defined by:
\begin{equation}\label{def: S}
\bs \ = \ \frac{\nabla \rho}{\rho} \ - \ \frac{\bg}{c^2} \ ,
\end{equation}
and can be shown to be proportional to the gradient of specific entropy.
Another parameter of interest in the fluid is the square of the \BV \ frequency $N^2$, which is related to
$\bs$ by:
\begin{equation}\label{def: N}
N^2 \ = \ \bs\cdot\bg \ = \ \frac{1}{\rho} \, \left( \nabla\rho \ - \ \frac{\rho}{c^2} \, \bg \right) \cdot\bg \ .
\end{equation}
The \BV \ frequency arises naturally when analyzing the local stability of the fluid since it provides
a simple way to formulate the Schwarzschild criterion \citep{S1906}.
An inspection of the expression of the energy reveals, indeed, that the local convective stability of the fluid is 
determined by the sign of $N^2$ (\eg \cite{FriSch78,bernard86}).
Actually, $N^2$ controls the non-seismic part of the spectrum of the elastic-gravitational operator,
$\sigma_e(\bcA)$:
\begin{equation}
\sigma_e(\bcA) \ = \ \left[
\Min(0,\iN2),\Max(0,\sN2)
\right],
\end{equation}
where $\iN2$ and $\sN2$ stand for the extrema of $N^2$ over $\earth_F$ \citep{bernard89}.
This implies that the corresponding squared eigenfrequencies range in the latter interval. 
In the Earth, these eigenfrequencies merely exceed $50 \, \mu$Hz,
a value which is well below that of the gravest seismic oscillation $_0S_2$.

In this paper, we only intend to compute the seismic part of the fluid outer core's response,
which is also affected by the variations of $N^2$.
Taking into account a fluid region within the framework 
of the finite element method is known to be a difficult problem, due to
the possible splitting of the zero eigenfrequency induced by the numerical discretization
of the elastic operator \citep{hamdi}. 
A key issue to produce a numerical solution free of spurious modes is the correct representation
of the null space of the elastic-gravitational operator, ${\cal N}(\bcA)$ \citep{br94}.
An alternative to the discretization of ${\cal N}(\bcA)$ is to solve the wave equation 
in the range of the operator, ${\cal R}(\bcA)$.
To proceed, we note from \eq (\ref{eq: motion_liq_u}) that an acceptable form for any displacement
field in ${\cal R}(\bcA)$ is:
\begin{equation}\label{eq: formu}
\bu \ = \ \nabla \chi \ + \  \xi \, \bs ,
\end{equation}
where $\chi$ and $\xi$ denote two arbitrary scalar fields.
Differentiating twice in time and identifying each term with the right-hand-side of 
\eq (\ref{eq: motion_liq_u}), we obtain two scalar wave equations, one for each potential:
\begin{eqnarray}
\ddot{\xi} & = &  c^2 \div \left(\nabla\chi + \xi \, \bs \right)
\ , \label{eq: motion_liq_xi} \\
\ddot{\chi} & = & \ddot{\xi} \ + \ \nabla\chi \, \cdot \bg \ + \ N^2 \, \xi  \ - \ \psi
\ . \label{eq: motion_liq_chi}
\end{eqnarray}
Eventually, the MRP $\psi$ appearing in \eqs (\ref{eq: motion_sol_b}) and (\ref{eq: motion_liq_chi}) 
is obtained by solving the Poisson-Laplace equation over the entire space. This writes:
\begin{equation}\label{eq: poisson}
\nabla^2 \psi  = \left\{
\begin{array}{ll}
 -4\pi G \, \divrhou & \mbox{in } \VS \ , \\[1mm]
 -4\pi G \, \div \left(\rho\,\nabla\chi +  \rho\,\xi \,\bs \right) & \mbox{in } \VF \ , \\
 0 & \mbox{outside } \earth\ ,  
\end{array}
\right.
\end{equation}
where $G$ is the gravitational constant.

\subsection{Boundary conditions}
The complete set of boundary conditions for displacement, traction and 
MRP can be found in \citet[][p.~104]{DahTro98}.
Here we recall these boundary conditions that concern the MRP or involve a solid-fluid interface. 

Let $\Sigma$ be a given interface in the medium.
The condition that the MRP must be continuous across $\Sigma$ reads:
\begin{equation}\label{eq: phicon}
\saut{\psi}_\Sigma \ = \ 0 \ , 
\end{equation}
where $\saut{ \quad }_\Sigma$ stands for the jump operator across $\Sigma$, defined in accordance
with the unit normal vector $\bnc$: $\saut{\psi}_\Sigma = \psi^+ - \psi^-$ and $\bnc$ points
from the $-$ to the $+$ side.
The normal derivative of $\psi$ can have a jump which is controlled by: 
\begin{equation}\label{eq: jump_dphidn}
\saut{\nabla\psi\cdot\bnc}_\Sigma \ = \ -4\pi G \,\saut{\rho\bu\cdot\bnc}_\Sigma \ .
\end{equation}
The condition that both traction and normal displacement must be continuous across the solid-fluid
boundaries writes as a set of equalities on $\SSF$:
\begin{eqnarray}
\bu\cdot\bnc & = & \left(\nabla\chi + \xi \, \bs\right) \cdot\bnc \, , \label{eq: uncon} \\
\bT(\bu)\cdot\bnc & = & \rho\,\ddot{\xi} \, \bnc \ . \label{eq: Tcon}
\end{eqnarray}
Note that to obtain \eq (\ref{eq: Tcon}) we have used \eqs (\ref{def: T_liq}), (\ref{eq: formu})
and (\ref{eq: motion_liq_xi}).
%
\subsection{Weak form}
%
The weak form of the wave equation in the solid regions is obtained after multiplying each side
of \eq (\ref{eq: motion_sol_a}) with an admissible displacement field $\bw$, then
integrating over $\VS$. This writes:
\begin{equation}\label{eq: wf_solid_a}
\left( \ddot{\bu} + \bcA(\bu) \, ; \, \rho\,\bw \right)_{\VS}
\ = \ 
\left( \bbf \, ; \, \bw \right)_{\VS}
\end{equation}
where $(\ ; \, )_{\VS}$ stands for the $\bL^2$ scalar product on $\VS$.
For example, integrating by parts the divergence of the stress tensor in \eq (\ref{eq: motion_sol_b})
yields:
\begin{equation}\label{eq: ipp0_divT}
- \left( \nabla\cdot \bT(\bu) 
\, ; \, \bw \right)_{\VS}
\ = \ 
\int_{\VS} \bT(\bu) \, \cdot \nabla\bw \ dV  \ - \
\int_{\SSF} \bT(\bu) \cdot\bnc \, \cdot \bw \ dS \ ,
\end{equation}
where $\bnc$ stands for the unit vector normal to $\SSF$ pointing away from $\VS$.
Note that the condition of free traction at the surface of the Earth \surf \ is naturally 
satisfied in \eq (\ref{eq: ipp0_divT}) as we have set the corresponding integral to zero.
On the contrary, the continuity of traction (\ref{eq: Tcon}) across the solid-fluid boundaries 
has to be enforced.
To proceed, we simply replace the traction vector in the surface integral
of \eq (\ref{eq: ipp0_divT}) with its fluid counterpart:
\begin{equation}\label{eq: ipp_divT}
- \left( \nabla\cdot \bT(\bu) 
\, ; \, \bw \right)_{\VS}
\ = \ 
\int_{\VS} \bT(\bu) \, \cdot \nabla\bw \ dV  \ - \
\int_{\SSF}  \rho \, \ddot{\xi} \, \bnc \, \cdot \bw \ dS \ .
\end{equation}
The weak form of the wave equation in the fluid regions is obtained similarly after dotting each side of
\eqs (\ref{eq: motion_liq_xi}) and (\ref{eq: motion_liq_chi}) with admissible potentials 
$\xit$ and $\chit$, integrating 
(possibly by parts) over $\VF$, then forcing the continuity of the normal displacement
(\ref{eq: uncon}) across the fluid-solid interfaces. One gets:
\begin{align}
\int_{\VF} \frac{1}{c^2} \, \ddot{\xi} \, \xit \ dV & = 
- \int_{\VF} \left(\nabla\chi + \xi \, \bs\right) \cdot \nabla\xit \ dV
+ \int_{\SSF} \!\!  \bu  \cdot \bnc \, \xit \ dS \ ,
\label{eq: wf_fluid_xi} \\
\intertext{and}
\int_{\VF} \frac{1}{c^2} \, \ddot{\chi} \, \chit \ dV & = \int_{\VF} 
\frac{1}{c^2} \,
\left(
\ddot{\xi} + \nabla\chi\cdot\bg + N^2\,\xi - \psi
\right) \, \chit \ dV \ . \label{eq: wf_fluid_chi}
\end{align}
In \eq (\ref{eq: wf_fluid_xi}), $\bnc$ denotes the unit vector normal to $\SSF$ that points outward
the fluid.
Note that the scaling factor $c^{-2}$ has been artificially included in \eq (\ref{eq: wf_fluid_chi}) 
in order to get the same left hand side as in \eq (\ref{eq: wf_fluid_xi}). This will make the description
of the time marching algorithm easier in section \ref{sec: specfem}.

Now, in order to establish the weak form of \eq (\ref{eq: poisson}), it is convenient to
first consider Poisson's equation within the finite (spherical) volume $\earthb$. 
Multiplying with an admissible potential $\psit$ defined over $\earthb$, then integrating by parts the 
Laplacian and the divergence we get:
\begin{align}
\int_{\earthb} \nabla\psi\cdot\nabla\psit \ dV 
& - \ \int_{\surfb} \nabla\psi\cdot\bnc \, \psit \ dS \ = \label{eq: fv_poisson0} \\
-4\pi G \left\{ \int_{\VS} \rho\,\bu \cdot \nabla\psit \ dV \right.
& - \int_{\surfb} \rho\,\bu \cdot \bnc \, \psit \ dS 
 + \left. \int_{\VF} \rho\,\left(\nabla\chi + \xi \, \bs\right) \cdot \nabla\psit \ dV \right\} \ , \nonumber 
\end{align}
with the boundary term involving the normal displacement being null, 
except in the absence of topography (\ie when $\surfb = \surf$).
It is important to note that the jump condition (\ref{eq: jump_dphidn}) 
across the solid-fluid interfaces is naturally taken into account in (\ref{eq: fv_poisson0}). 
This property, which stems from the potential decomposition (\ref{eq: formu}), is a key
argument that guided our choice to work with the displacement field (and not the velocity) in the fluid.
%
\subsection{$DtN$ operator}
%
The harmonic behaviour of $\psi$ outside $\earthb$ has not been considered yet.
In order to proceed, let $\phiint$ denote the MRP interior to $\earthb$.
At the (spherical) surface $\surfb$, consider the expansion of $\phiint$ onto the 
orthonormal basis of real spherical harmonics $\Ylm$ (see \cite{DahTro98}, p.851):
\begin{equation}
\phiint (b,\theta,\varphi) \ = \ \sum_{l=0}^{\infty} \sum_{m=-l}^{l} \phiint^{\,lm}(b) \, \Ylm(\theta,\varphi) \ ,
\end{equation}
where $(\theta,\varphi)$ are the spherical coordinates and
where $\phiint^{\,lm}(b) = \int_{\surfb} \phiint \Ylm \ dS$.
It is straightforward to extend $\phiint$ continuously to a potential $\phiext$
that satisfies Laplace's equation outside $\earthb$ and vanishes at infinity:
\begin{equation}\label{eq: phi_ext}
\phiext (r,\theta,\varphi) \ = \ \sum_{l=0}^{\infty} \sum_{m=-l}^{l} \phiint^{\,lm}(b) \,
\left(\frac{b}{r}\right)^{l+1} \, \Ylm(\theta,\varphi) \ , \quad r \ge b \ .
\end{equation}
The normal derivative of $\phiext$ on $\surfb$ is readily obtained by differentiating the previous 
expression with respect to $r$ :
\begin{equation}\label{eq: DtN}
\nabla\phiext\!\cdot\bnc \, (b,\theta,\varphi) \ = \
- \frac{1}{b} \sum_{l=0}^{\infty} (l+1) \sum_{m=-l}^{l} \, \phiint^{\,lm}(b) \,
\Ylm(\theta,\varphi) \ .
\end{equation}
\Eq (\ref{eq: DtN}) which relates the normal derivative of the potential to the potential
itself is called a Dirichlet-to-Neumann ($DtN$) operator on the spherical boundary $\surfb$.
Its action, which is non-local, is rather simple to express in the spherical harmonics basis: it consists in 
multiplying each coefficient with $\frac{-l-1}{b}$.
Recall that the condition that the normal derivative of a given field is proportional to the 
field at the surface is referred to as a Robin boundary condition. 
Applying the $DtN$ operator is therefore equivalent to imposing a Robin boundary condition on 
every component of the spherical harmonics expansion of the original potential, and this yields
a well-posed problem.

Taking into account the jump condition (\ref{eq: jump_dphidn}) across $\surfb$, we can 
write the final weak form of the Poisson-Laplace equation as:
\begin{align}
- \int_{\earthb} \nabla\psi\cdot\nabla\psit \ dV 
& + \ \int_{\surfb} \nabla\phiext\!\cdot\bnc \, \psit \ dS \ = \label{eq: fv_poisson} \\
4\pi G \left\{ \int_{\VS} \rho\,\bu \cdot \nabla\psit \ dV \right.
& \left.  + \int_{\VF} \rho\,\left(\nabla\chi + \xi \, \bs\right) \cdot \nabla\psit \ dV \right\} \ , \nonumber
\end{align}
with:
\begin{equation}\label{eq: corr_surf}
\int_{\surfb} \nabla\phiext\!\cdot\bnc \, \psit \ dS \ = \
-\frac{1}{b} \sum_{l=0}^{\infty} (l+1) \sum_{m=-l}^{l} \, \phiint^{\,lm}(b) \, \psit^{\,lm}(b) \ .
\end{equation}
In practice, the infinite sum present in \eq (\ref{eq: corr_surf}) will be limited to angular orders
$l<\lmax \ $.
Note that the effect of the truncation is to apply a Neumann boundary condition to 
the high wavenumber content of the MRP, which according to \eq (\ref{eq: phi_ext}) 
is asymptotically consistent with the behaviour of the MRP outside $\earthb$.

\section{Numerical approximation}\label{sec: specfem}
This section deals with the numerical approximation of the wave equation in a self-gravitating Earth, 
which we achieve in two steps.
First, the SEM is applied to the weak form of the equations 
in the space domain. Then a finite difference scheme is used to advance the system in time.

For the sake of conciseness, details of the method are avoided 
as much as possible unless this prevents the paper from being self-contained.
The reader is referred to \citep{kv98} and to \citep{kt99} for a general 
description of the SEM applied to the elastic wave equation, and to 
(\citeauthor{KomTro02a} \citeyear{KomTro02a}; \citeyear{KomTro02b}) and
\citep{ChaCapVil03} for its extension to global seismology, 
including its parallel implementation on modern computers with distributed memory.

\subsection{Spatial discretization}
\subsubsection{Hexahedral Mesh}
The first discretization step consists in decomposing the spherical Earth into a collection of
non-overlapping hexahedral elements. 
This process is detailed in \citep{ChaCapVil03}, where non-conforming interfaces are introduced
to avoid an artificial refinement of the grid with depth. 
Such a strategy allows the refinement (or coarsening) of the mesh to be spatially localized, 
the complexity being related to the continuity requirements between elements 
that do not match across the interfaces.
For the sake of simplicity, this paper is restricted to the case of a spherical,
geometrically conforming mesh such as the one represented in fig.~\ref{fig: mesh}.
Note that taking into account the elliptical figure of the Earth or accounting for surface topography
would require in the self-gravitating case to extend the mesh outward the artificial boundary $\surfb$. 

\subsubsection{Spectral element method}
Based upon the 3D tiling of the sphere, the MRP ($\psi$) as well as the displacement in the solid ($\bu$)
and the potentials in the fluid ($\chi$ and $\xi$) are approximated using continuous 
tensorized polynomials.
Note that the continuity of the normal displacement within the fluid regions is naturally
satisfied in the weak forms (\ref{eq: wf_fluid_xi}) and (\ref{eq: wf_fluid_chi}).

The basis of polynomials used on each spectral element are defined as the shape functions of the 
collocation points. 
One of the particularity of the SEM is that the collocation points are the so-called 
Gauss-Lobatto-Legendre points, \ie the exact same points that are used to evaluate
the integrals present in the weak form of the equations.
One consequence of this choice is that the matrix representation of the $L^2$ scalar product is diagonal, 
a property that allows to design explicit time schemes (see \eg \cite{kv98} and \cite{kt99}).
\subsection{Time evolution}
The different steps of the spatial discretization yield a system of ordinary differential 
equations in time, which writes:
\begin{alignat}{4}
\bM_S \, \ddot{\bd}(t) & + \ \bK_S \, \bd (t)  \ + \ \bG \, \bPhi (t) &  + \ \bC_{SF}\,\bddxi(t)& \ = \ \bF (t) \label{eq: ODEa}\\
\bM_F \, \bddxi(t) & + \ \bK_F \, \left(\bxi,\bchi\right)(t)  & +  \ \bC_{FS}\, \bd (t) & \ = \ \mathbf{0} \label{eq: ODEb}\\
\bM_F \, \bddchi(t) & + \ \bB_F \, \left(\bddxi,\bxi,\bchi,\bPhi \right)(t)&& \ = \  \mathbf{0} \label{eq: ODEc}\\
&& \bP \, \bPhi(t) & \ = \  \bD \, \left(\bd,\bxi,\bchi\right) (t) \label{eq: ODEd}
\end{alignat}
In the previous equations, $\bd$ stands for the displacement vector in the solid regions,
$\bF$ is the approximation of the source term and $\bPhi$, $\bchi$, $\bxi$
respectively denote the nodal values of the MRP and of the displacement potentials 
in the fluid.
$\bM_S$ is the mass matrix in the solid regions, \ie the matrix representation of
the $L^2$ scalar product weighted by density.
Similarly, $\bM_F$ is the matrix representation of the scalar product in the fluid regions
weighted by the quantity $c^{-2}$.
As outlined before, both matrices are diagonal.
$\bK_S$ and $\bK_F$ are the stiffness matrices which arise from the approximation of the
volume integrals in \eqs (\ref{eq: ipp_divT}) and (\ref{eq: wf_fluid_xi}).
The discretization of the surface integrals in the latter equations yields the solid-fluid 
coupling matrices $\bC_{SF}$ and $\bC_{FS}$.
$\bB_F$ arises from the discretization of the right hand side of 
\eq (\ref{eq: wf_fluid_chi}) and only involves a
pointwise operation on $\bddxi$, $\bxi$, $\nabla\bchi$ and $\bPhi$.
Finally, $\bG$, $\bD$ and $\bP$ are the matrix representations of the gradient, divergence
and Poisson-Laplace operator, respectively. 
Note that $\bD$ contains the factor $4\pi G \rho$ and that
$\bP$ is symmetric according to \eqs (\ref{eq: fv_poisson}) and (\ref{eq: corr_surf}).

To advance the equations forward in time we use the explicit, second-order accurate, Newmark 
scheme \citep[\eg][]{hugues87}.
Let for example $\bX_n$ denote the snapshot at time $t_n$ of one of the unknown vectors 
$\bd$, $\bchi$ or $\bxi$ involved in \eqs (\ref{eq: ODEa}--\ref{eq: ODEc}).
The values of $\bX$ and its time derivative at the next time step are extrapolated as follows:
\begin{eqnarray}
\bX_{n+1} & = & \bX_n + \Delta t \,\dot{\bX}_n + \frac{\Delta t^2}{2} \,\ddot{\bX}_n 
\label{eq: FDa}\\
\dot{\bX}_{n+1} & = & \dot{\bX}_n + \frac{\Delta t}{2} \left( \ddot{\bX}_n + \ddot{\bX}_{n+1} \right)
\label{eq: FDb}
\end{eqnarray}
As it is readily seen from the previous equations, the algorithm is fully explicit in terms of $\bX$ and
consists in a simple centered finite difference scheme in $\dot{\bX}$.
The process of updating the time derivatives of $\bX$ is achieved in two steps: 
first $\ddot{\bX}_{n+1}$ is computed from the discrete version
of the wave equation (\ref{eq: ODEa}--\ref{eq: ODEc}) by inverting a diagonal mass matrix ($\bM_S$ or $\bM_F$), 
then $\dot{\bX}_{n+1}$ can be updated using (\ref{eq: FDb}).
Note that the wave equation has to be solved in the fluid regions first, 
since the coupling operator $\bC_{SF}$ in \eq (\ref{eq: ODEa}) acts on 
$\bddxi_{n+1}$ which is not known at time $t_n$.

Let us stress that the coupling between the fluid and the solid regions does
not require iterations of \eqs (\ref{eq: FDa},\ref{eq: FDb}) as this would be the case 
if a velocity potential formulation was used \citep[\eg][]{kbt00a,ChaCapVil03}.
This attractive property stems from the potential decomposition (\ref{eq: formu}) 
applied to the displacement which is the explicit variable in the Newmark scheme.

The previous remark remains valid when the full effects of self-gravitation are taken into account.
The computation of the MRP from the displacement field is indeed explicit in the sense that it 
does not involve any time derivative $\dot{\bX}$ or $\ddot{\bX}$.
Needless to say, this task is expensive as it requires to formally invert the 
symmetric, ill-conditioned matrix $\bP$ \citep[\eg][]{DeFiMu02}.
In practice, we solve \eq (\ref{eq: ODEc}) for the MRP with a conjugate gradient (CG) 
method which iterations are stopped when the residual is decreased by a factor $\epsilon$ to be chosen. 
The issue of building an efficient preconditioner for the Poisson-Laplace solver is not addressed
in this paper, but it is certainly critical in order to avoid a performance bottleneck.

\section{Numerical results}\label{sec: results}

In this section, we demonstrate the validity of our approach through a couple of examples
for which a reference, semi-analytical, solution can be derived.
First, the two potentials formulation is tested within the Cowling approximation,
\ie without computing the MRP, for models having a constant \BV \ frequency.
Then, the effects of mass redistribution are included in a simplified version of the PREM model \citep{prem}.

\subsection{Validation of the two-potentials formulation}

\begin{figure}
\psfrag{Rho}{$\rho$}
\centerline{\includegraphics
[width=.5\linewidth,angle=90]
{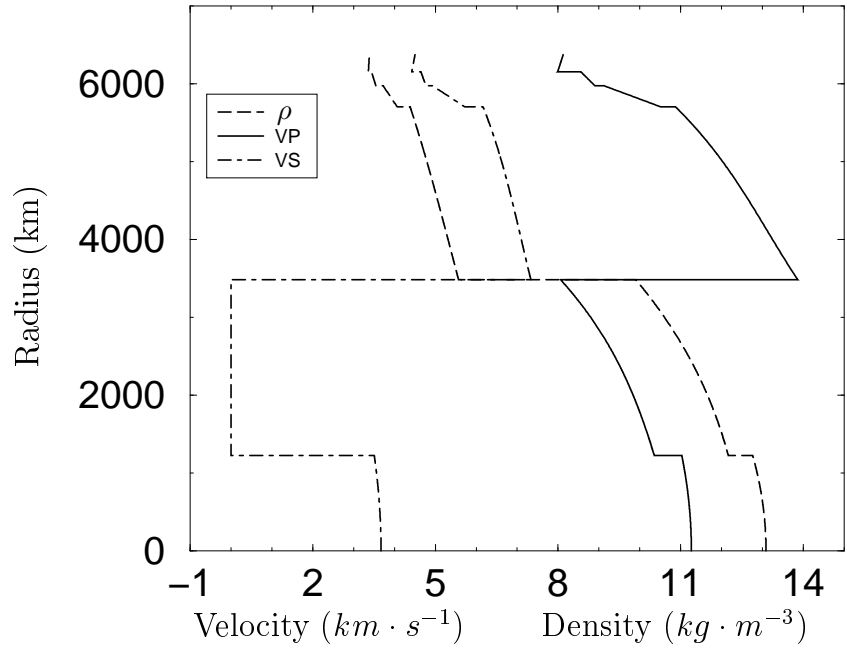}}
\caption{
Variation with depth of density (dashed curve), $P$-velocity (solid curve) 
and $S$-velocity (dot-dashed curve) within the Earth-like model used in this paper.
The model is adapted from PREM \citep{prem} with the complexity of the lithospheric structure 
being removed to simplify computation.
}
\label{fig: PREL}
\end{figure}

In order to define some benchmarks to test our formulation, we consider the radial Earth model
of fig.~\ref{fig: PREL}.
The model is adapted from PREM, with a smaller number of regions 
(6 instead of 13). In particular, the details of the crustal structure as well as the presence of
a global ocean are ignored to ease the computation.
This reference model is further constrained to fit a given profile of the squared \BV \ 
frequency in the fluid outer core.
To proceed, we simply vary the $P$-velocity in \eq (\ref{def: N}),
keeping the density, its gradient and the gravitational acceleration unchanged.
Note that a realistic way would be to adjust density rather than $P$-velocity
(see \eg \cite{WuRoc93}) because the latter is much better constrained in the Earth. 
However, acting on the $P$-velocity profile is straightforward and still fully acceptable for numerical
validation purposes.

\begin{figure}
\centerline{\includegraphics
[width=.5\linewidth,angle=90]
{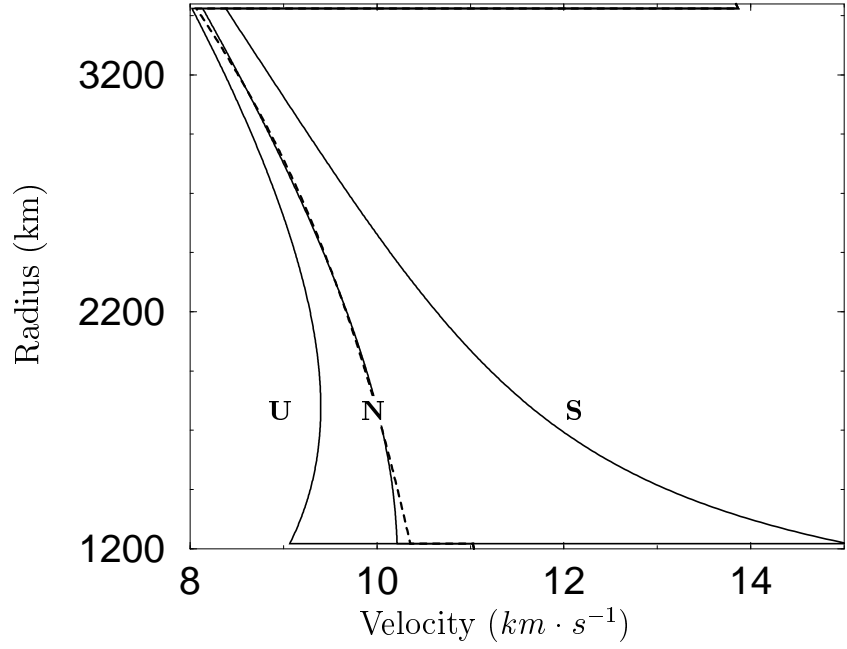}}
\caption{
Different profiles of $P$-velocity used to test the two-potentials formulation in the fluid outer core.
The dashed curve represents the variation of the sound speed within the model detailed in
fig.~\ref{fig: PREL}.
Each solid curve corresponds to a modification of that profile such that the square of the \BV \ frequency 
is constant throughout the fluid.
The label `N' corresponds to a neutrally stratified outer core, whereas `S' (resp.~`U') 
stands for a stable (resp.~unstable) stratification for which $N^2=10^{-7} \ \rad^2\cdot \s^{-2}$
(resp.~$N^2=-5\,10^{-8} \ \rad^2\cdot \s^{-2}$).
}
\label{fig: SUN}
\end{figure}

Fig.~\ref{fig: SUN} shows three models that were built following the above procedure.
The $\mbox{`N'}$ label refers to a neutrally stratified outer core (\ie with $N^2 = 0$), whereas
the models labelled `S' and `U' correspond to a stable and unstable stratification, respectively.
For the sake of simplicity, we chose the value of the squared \BV \ frequency to be constant 
throughout models 'S' and 'U', respectively equal to $N^2=10^{-7} \ \rad^2\cdot \s^{-2}$ and 
$N^2=-5\,10^{-8} \ \rad^2\cdot \s^{-2}$.
These values correspond to the extrema that are expected from the inversion of seismic free oscillations
of the Earth (\cite{Masters79}, Valette \& Lesage, unpublished).
Note that the values of $N^2$ within PREM are about one order of magnitude smaller, as illustrated
by the similarity of the PREM $P$-velocities to those of a neutrally stratified profile.

\begin{figure}
\centerline{\includegraphics
[width=.5\linewidth,angle=90]
{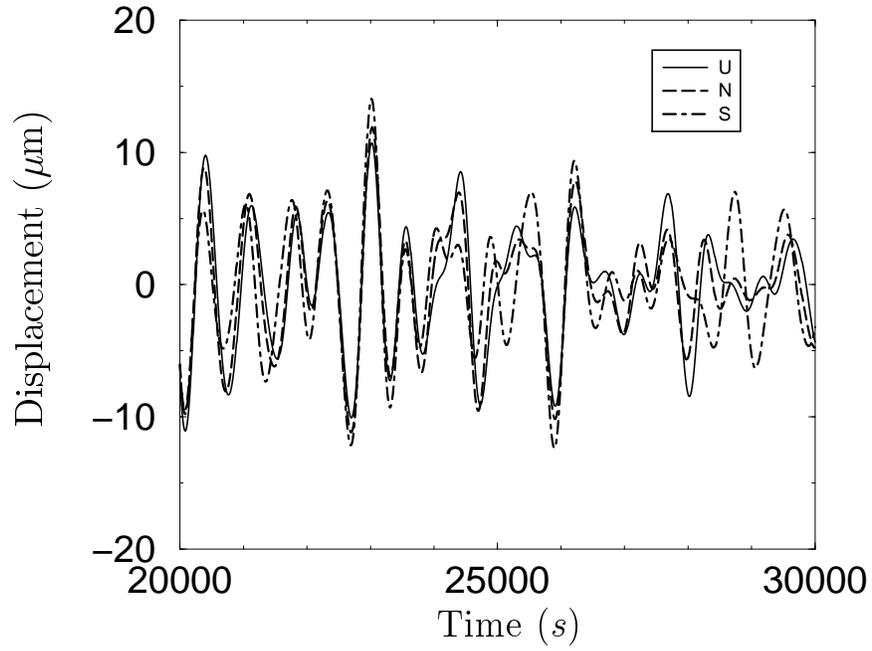}}
\caption{
Time window of the longitudinal surface displacement recorded at $90 \deg$ in the models labelled
`N', `S' and `U' in fig.~\ref{fig: SUN}.
The large waveform differences stem from the sensitivity of the seismic modes to the variation of
the $P$-velocity within the three models.
}
\label{fig: comp_SUN}
\end{figure}

\begin{figure}
\centerline{\includegraphics
[width=.8\linewidth,angle=90]
{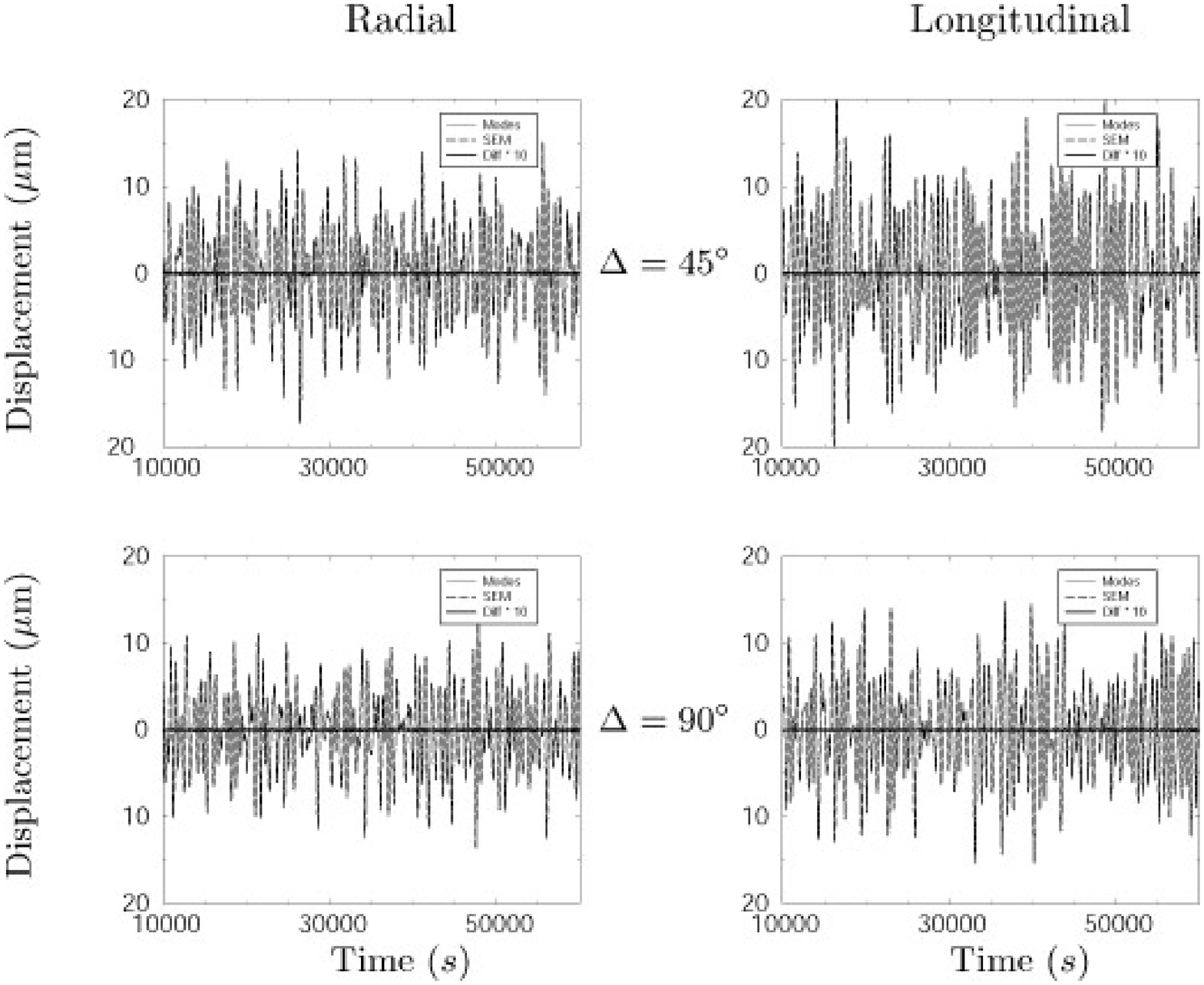}}
\caption{
Radial (left panel) and longitudinal (right panel) components of the surface displacement 
recorded at  $45 \deg$ (top) and $90 \deg$ (bottom) in the model labelled `S' in fig.~\ref{fig: SUN}.
In each plot, the spectral element solution (dashed line) is compared to the normal modes reference 
(solid thin line) and the residual (solid bold line) is amplified by a factor of 10. 
}
\label{fig: results_S}
\end{figure}

\begin{figure}
\centerline{\includegraphics
[width=.8\linewidth,angle=90]
{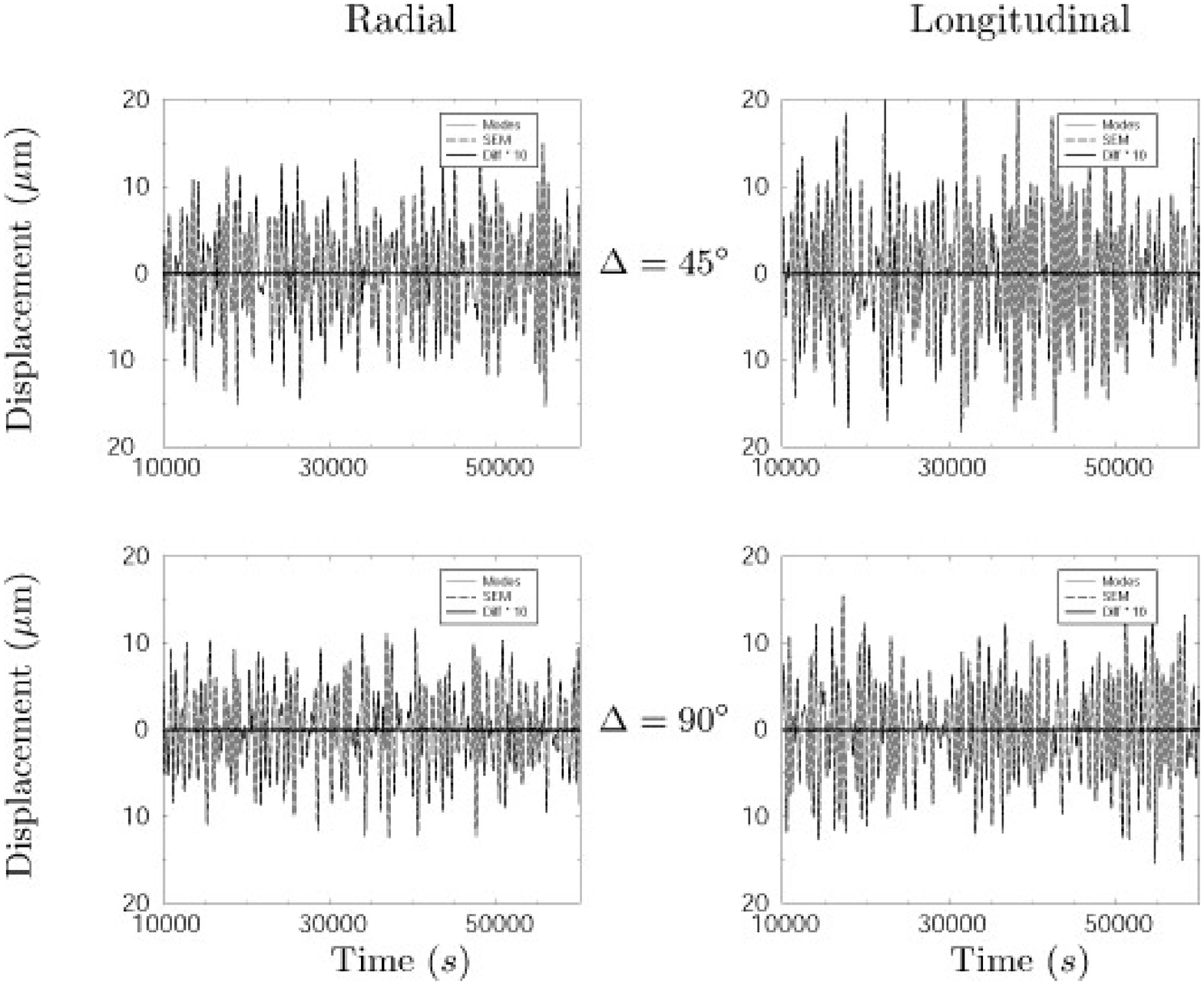}}
\caption{
Radial (left panel) and longitudinal (right panel) components of the surface displacement 
recorded at  $45 \deg$ (top) and $90 \deg$ (bottom) in the model labelled `U' in fig.~\ref{fig: SUN}.
In each plot, the spectral element solution (dashed line) is compared to the normal modes reference 
(solid thin line) and the residual (solid bold line) is amplified by a factor of 10. 
}
\label{fig: results_U}
\end{figure}

\begin{figure}
\centerline{\includegraphics
[width=.5\linewidth,angle=0]
{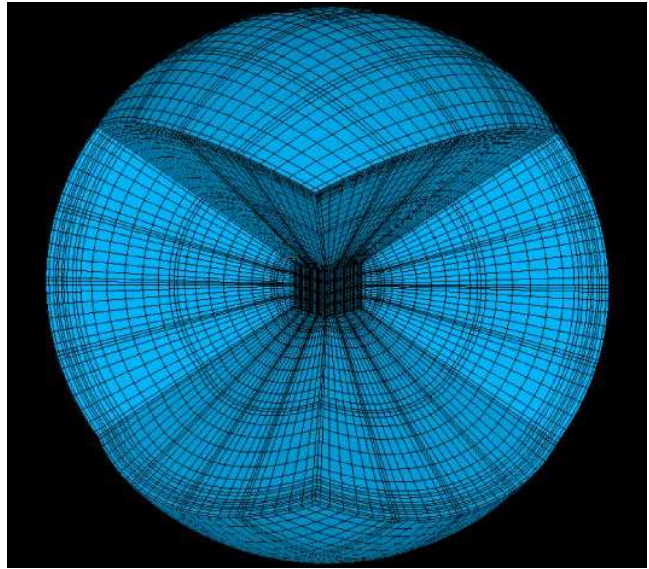}}
\caption{
Spectral element mesh used to compute the results shown in figs.~\ref{fig: results_S} and 
\ref{fig: results_U}.
Two blocks of the 3D mesh have been removed to allow a view inside the volume. 
The mesh is composed of 640 spectral elements with varying polynomial order, for a total
number of gridpoints equal to 334,368.
The process of building the mesh is detailed in \citep{ChaCapVil03}.
This image was generated using the visualization software pV3 ({\tt http://raphael.mit.edu/pv3/pv3.html}).
}
\label{fig: mesh}
\end{figure}

All three models are excited by a shallow explosive point source which time dependence is a Ricker wavelet 
(\ie the second derivative of a Gaussian bell) with dominant frequency $f_0 = 1\, \mbox{mHz}$.
The source is located at one grid-point from the Equator, at latitude $\theta_s \simeq -1.128^\circ$
and depth $d_s \simeq 61$ km, 
and the receivers sit along the Equator.
Fig.~\ref{fig: comp_SUN} shows the longitudinal displacement recorded at an epicentral distance of
$90^\circ$ in the three models. 
The traces were computed within the Cowling approximation using a summation 
of the eigenmodes of each model.
The waveform differences illustrate the sensitivity of the seismic waves to the stratification of the 
fluid core and suggest that models `U' and `S' constitute a demanding benchmark for
the two potentials formulation.
In figs.~\ref{fig: results_S} and \ref{fig: results_U}, the spectral element results obtained in those 
two models are compared to the modal solutions for a couple of epicentral distances. 
The two solutions are in very close agreement with the largest relative differences being as small as one
per mil over the time interval considered.

The spectral element grid used to carry out the calculations is shown in fig.~\ref{fig: mesh}. 
It consists of 640 elements in which the polynomial degree varies from 3 to 10 in the radial direction
and is kept constant, equal to 8, in the tangential direction. The total number of gridpoints is
334,368 corresponding to a number of points per wavelength much greater than 5, which is the empirical ratio
to get an accurate solution (\eg \cite{kv98}). 
This explains the perfect match between the spectral element calculations and the reference solutions.

\subsection{Validation of the whole formulation}
As a last example, we consider the computation of the elastic-gravitational response of the Earth model of 
fig.~\ref{fig: PREL}. 
This test presents all the difficulties mentioned in this paper: the stratification of the
fluid core is arbitrary and the physical description includes the full effects of self-gravitation. 

The parameters of the simulations are slightly different than above, since the source dominant frequency
is set to a graver value $f_0= 0.5\, \mbox{mHz}$, and the source latitude is now 
$\theta_s \simeq -2.64^\circ$.
The spectral element grid is consequently adapted, and roughly coarsened by a factor of two in each 
direction compared to the one of fig.~\ref{fig: mesh}.

\begin{figure}
\centerline{\includegraphics
[width=.5\linewidth,angle=90]
{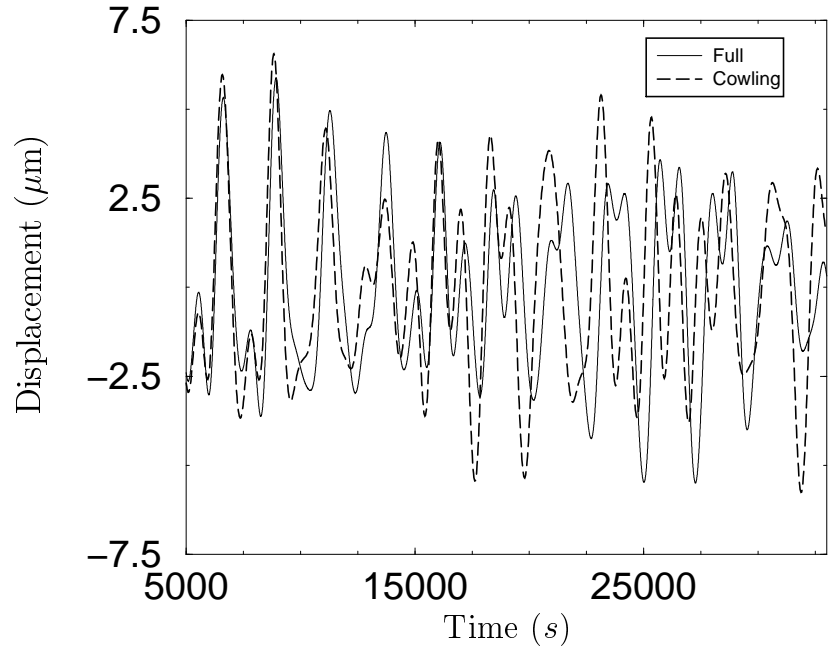}}
\caption{
Longitudinal component of the surface displacement recorded at $90 \deg$ in the Earth model of
fig.~\ref{fig: PREL}. 
The trace computed with the full treatment of self-gravitation (solid thin line)
is compared to the one computed within the Cowling approximation (dashed bold line).
The waveform differences illustrate that the effect of the MRP cannot be 
neglected at the frequencies considered in this experiment.
}
\label{fig: comp_PREL}
\end{figure}

In order to check that the test is demanding enough with regard to the implementation of self-gravitation, 
we compare in fig.~\ref{fig: comp_PREL} the surface longitudinal displacement recorded with or without
including the perturbation of the gravitational potential.
Both traces were computed by normal modes summation and recorded at an epicentral distance of $90^\circ$
for about 10 hours.
The differences in phase and amplitude illustrate that the Cowling approximation is not valid in the
frequency range of the experiment.

\begin{figure}
\centerline{\includegraphics
[width=.5\linewidth,angle=90]
{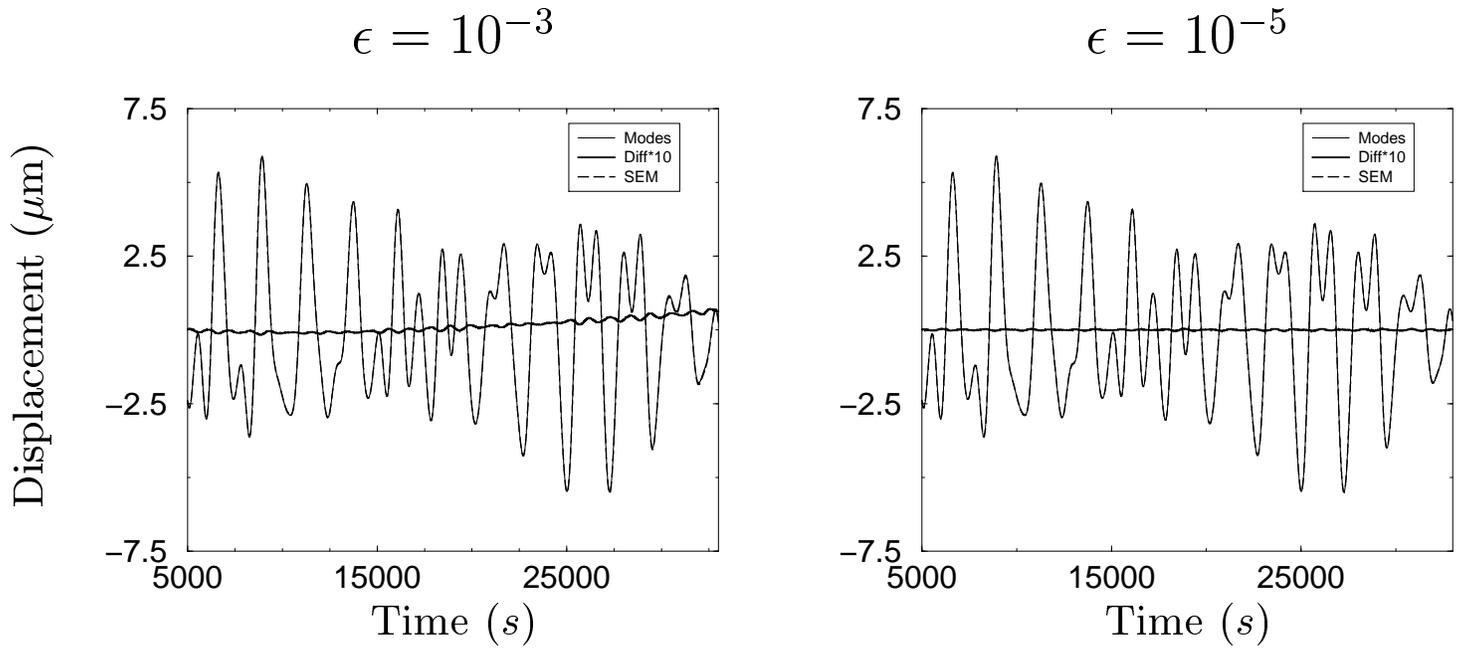}}
\caption{
Longitudinal surface displacements recorded at $90 \deg$ in the Earth model of 
fig.~\ref{fig: PREL}.
The left (resp. right) plot corresponds to a low (resp. high) accuracy test in which the 
CG iterations used to compute the MRP are stopped when the residual is decreased 
by 3 (resp. 5) orders of magnitude.
In each plot, the spectral element solution (dashed line) is compared to the normal modes reference 
(solid thin line) and the residual (solid bold line) is amplified by a factor of 10.
}
\label{fig: results_PREL}
\end{figure}

Finally, the results obtained with the SEM are compared to the reference solution in 
fig.~\ref{fig: results_PREL}.
Two cases are considered that correspond to a different accuracy of the spectral element solution regarding 
the CG resolution of the discrete Poisson-Laplace equation (\ref{eq: ODEc}). 
In the first case the CG iterations are stopped when the residual is decreased by three orders of
magnitude, which means that $\epsilon=10^{-3}$. 
The resulting spectral element solution is clearly not accurate enough and contains a secular term
that seems to break the conservation of energy at the discrete level.
To correct this behaviour, we consider a second test where the stopping criterion is fixed to 
$\epsilon=10^{-5}$. In that case, the calculation is stable upon the time interval considered and the
accuracy of the spectral element solution is found to be acceptable, its relative difference with the
reference solution being less than a few per mil.

In each of the previous cases, the angular order truncation in \eq (\ref{eq: corr_surf}) was set
to $\lmax =20$, based on the {\em a priori} knowledge of the dispersion relation in PREM.
The effect of underestimating the truncation order is to add oscillations to the spectral
element solution (not shown in this paper). 
It is interesting to note that the two possible sources of numerical errors
($\epsilon$ too big or $\lmax$ too small) lead to a different signature. 
This provides two different diagnostics that permit to build a spectral element solution with
arbitrary accuracy.

\section{Conclusions}
We have shown how the SEM should be adapted to account for two effects 
relevant to global seismology: the full treatment of self-gravitation and
the ability to consider any density stratification in the fluid outer core.
The accuracy of the method has been illustrated through a series of
numerical tests conducted in spherically symmetric models.
With the incorporation of the two aforementioned effects, we believe the SEM will provide 
new estimates of the elastic-gravitational response of 3D models of the Earth.

\begin{acknowledgments}
E.~C.~ greatly acknowledges the numerous discussions he had with the members of the seismological task force 
in Princeton University, USA, where this work was initiated.
The writing of the manuscript has benefited from the careful reading of Ludovic Margerin and 
Alexandre Fournier.
The computations presented in this paper were performed both 
at the Service Commun de Calcul Intensif (SCCI) at the Observatory in Grenoble, France and 
at the Centre Informatique National de l'Enseignement Sup\'erieur (CINES) in Montpellier, France.
\end{acknowledgments}

%

\bibliography{\bibPATH/biblio}

\begin{thebibliography}{25}
\expandafter\ifx\csname natexlab\endcsname\relax\def\natexlab#1{#1}\fi

\bibitem[Berm\'udez \& {Rodr\'\i guez}(1994)]{br94}
Berm\'udez, A. \& {Rodr\'\i guez}, R., 1994, Finite element computation of the
  vibration modes of a fluid-solid system, {\it Comput. Methods Appl. Mech.
  Engrg.\/}, {\bf 119}, 355--370.

\bibitem[Bettess(1992)]{Bet92}
Bettess, P., 1992, {\it Infinite {E}lements\/}, Penshaw Press, Sunderland,
  England.

\bibitem[Capdeville et~al.(2003)Capdeville, Chaljub, Vilotte, \&
  Montagner]{Capetal03}
Capdeville, Y., Chaljub, E., Vilotte, J.-P., \& Montagner, J.-P., 2003,
  Coupling the spectral element method with a modal solution for elastic wave
  propagation in global earth models, {\it Geophys. J. Int.\/}, {\bf 152 (1)},
  34--67.

\bibitem[Chaljub(2000)]{Cha00}
Chaljub, E., 2000, {\it Mod\'elisation num\'erique de la propagation d'ondes
  sismiques en g\'eom\'etrie sph\'erique : application \`a la sismologie
  globale ({N}umerical modeling of seismic wave propagation in spherical
  geometry: application to global seismology)\/}, Ph.{D.} thesis, Universit\'e
  Paris {VII}, Denis Diderot.

\bibitem[Chaljub et~al.(2003)Chaljub, Capdeville, \& Vilotte]{ChaCapVil03}
Chaljub, E., Capdeville, Y., \& Vilotte, J.-P., 2003, Solving elastodynamics in
  a fluid-solid heterogeneous sphere: a parallel spectral element approximation
  on non-conforming grids, {\it J. Comput. Phys.\/}, {\bf 187 (2)}, 457--491.

\bibitem[Cowling(1941)]{Cow41}
Cowling, T.~G., 1941, The non-radial oscillations of polytropic stars, {\it
  Mon. Not. Roy. Astron. Soc.\/}, {\bf 101}, 369--373.

\bibitem[Dahlen \& Tromp(1998)]{DahTro98}
Dahlen, F.~A. \& Tromp, J., 1998, {\it Theoretical {G}lobal {S}eismology\/},
  Princeton University Press, Princeton, NJ.

\bibitem[Deville et~al.(2002)Deville, Fischer, \& Mund]{DeFiMu02}
Deville, M.~O., Fischer, P.~F., \& Mund, E.~H., 2002, {\it High-Order Methods
  for Incompressible Fluid Flow\/}, Cambridge University Press, Cambridge, UK.

\bibitem[Dziewonski \& Anderson(1981)]{prem}
Dziewonski, A.~M. \& Anderson, D.~L., 1981, Preliminary {R}eference {E}arth
  {M}odel, {\it Phys. Earth Planet. Int.\/}, {\bf 25}, 297--356.

\bibitem[Friedman \& Schutz(1978)]{FriSch78}
Friedman, J.~L. \& Schutz, B.~F., 1978, Secular instability of rotating
  {N}ewtonian stars, {\it Ap J.\/}, {\bf 221}, 937--957.

\bibitem[Gerdes \& Demkowicz(1996)]{GerDem96}
Gerdes, K. \& Demkowicz, L., 1996, Solution of 3{D}-{L}aplace and {H}elmholtz
  equations in exterior domains using hp-infinite elements, {\it Comput.
  Methods Appl. Mech. Engrg.\/}, {\bf 137}, 239--273.

\bibitem[Givoli(1992)]{Giv92}
Givoli, D., 1992, {\it Numerical Methods for Problems in Infinite Domains\/},
  Elsevier Science Publishers, Amsterdam.

\bibitem[Hamdi et~al.(1978)Hamdi, Ousset, \& Verchery]{hamdi}
Hamdi, M., Ousset, Y., \& Verchery, G., 1978, A displacement method for the
  analysis of vibrations of coupled fluid-structure systems, {\it Int. J. Num.
  Meth. Engrg.\/}, {\bf 13}, 139--150.

\bibitem[Hugues(1987)]{hugues87}
Hugues, T. J.~R., 1987, {\it The finite element method, linear static and
  dynamic finite element analysis\/}, Prentice-Hall International.

\bibitem[Komatitsch \& Tromp(1999)]{kt99}
Komatitsch, D. \& Tromp, J., 1999, Introduction to the spectral element method
  for three-dimensional seismic wave propagation, {\it Geophys. J. Int.\/},
  {\bf 139}, 806--822.

\bibitem[Komatitsch \& Tromp(2002{\natexlab{a}})]{KomTro02a}
Komatitsch, D. \& Tromp, J., 2002, Spectral-element simulations of global
  seismic wave propagation, part {I}: {V}alidation, {\it Geophys. J. Int.\/},
  {\bf 149}, 390--412.

\bibitem[Komatitsch \& Tromp(2002{\natexlab{b}})]{KomTro02b}
Komatitsch, D. \& Tromp, J., 2002, Spectral-element simulations of global
  seismic wave propagation, part {II}: 3-{D} models, oceans, rotation, and
  gravity, {\it Geophys. J. Int.\/}, {\bf 150}, 303--318.

\bibitem[Komatitsch \& Vilotte(1998)]{kv98}
Komatitsch, D. \& Vilotte, J.-P., 1998, The spectral element method: an
  effective tool to simulate the seismic response of 2{D} and 3{D} geological
  structures, {\it Bull. Seismol. Soc. Am.\/}, {\bf 88}, 368--392.

\bibitem[Komatitsch et~al.(2000)Komatitsch, Barnes, \& Tromp]{kbt00a}
Komatitsch, D., Barnes, C., \& Tromp, J., 2000, Wave propagation near a
  fluid-solid interface: a spectral element approach, {\it Geophysics\/}, {\bf
  65 (2)}, 623--631.

\bibitem[Masters(1979)]{Masters79}
Masters, G., 1979, Observational constraints on the chemical and thermal
  structure of the {E}arth's interior, {\it Geophys. J. R. Astron. Soc.\/},
  {\bf 57}, 507--534.

\bibitem[Schwarzschild(1906)]{S1906}
Schwarzschild, K., 1906, \"{U}ber das {G}leichgewicht der {S}onneatmosph\"{a}re
  ({O}n the equilibrium of the {S}un's atmosphere), {\it G\"ottingen
  Nachrichten\/}, {\bf 1}, 41.

\bibitem[Valette(1986)]{bernard86}
Valette, B., 1986, About the influence of pre-stress upon the adiabatic
  perturbations of the {E}arth, {\it Geophys. J. Roy. Astron. Soc.\/}, {\bf
  85}, 179--208.

\bibitem[Valette(1989)]{bernard89}
Valette, B., 1989, Spectre des vibrations propres d'un corps \'elastique,
  auto-gravitant, en rotation uniforme et contenant une partie fluide ({F}ree
  oscillations spectrum of an elastic, self-gravitating, uniformly rotating
  body with a fluid inclusion), {\it C. R. Acad. Sci. Paris\/}, {\bf 309, I},
  419--422.

\bibitem[Wu \& Rochester(1990)]{WuRoc90}
Wu, W.-J. \& Rochester, M.~G., 1990, Core dynamics: the two-potential
  description and a new variational principle, {\it Geophys. J. Int.\/}, {\bf
  103}, 697--706.

\bibitem[Wu \& Rochester(1993)]{WuRoc93}
Wu, W.-J. \& Rochester, M.~G., 1993, Computing core oscillation eigenperiods
  for the rotating {Earth}: a test of the subseismic approximation, {\it Phys.
  Earth Planet. Inter.\/}, {\bf 78}, 33--50.

\end{thebibliography}

\newpage

\clearpage

\clearpage

\clearpage

\clearpage

\clearpage

\clearpage

\clearpage

\end{document}